\documentclass[preprint,aps]{revtex4}
\usepackage{amssymb,epsf}

\begin{document}

\title{Thermodynamics of Rotating Black Branes in \\
Gauss-Bonnet-nonlinear Maxwell Gravity}
\author{S. H. Hendi$^{1,2,3}$\footnote{email address: hendi@mail.yu.ac.ir} and B. Eslam Panah$^{1}$}
\affiliation{$^1$Physics Department, College of Sciences, Yasouj University, Yasouj
75914, Iran\\
$^2$Research Institute for Astrophysics and Astronomy of Maragha (RIAAM),
P.O. Box 55134-441, Maragha, Iran\\
$^3$National Elite Foundation, P.O. Box 19615-334, Tehran, Iran}

\begin{abstract}
We consider the Gauss-Bonnet gravity in the presence of a new class of
nonlinear electromagnetic field, namely, power Maxwell invariant. By use of
a suitable transformation, we obtain a class of real rotating solutions with
$k$ rotation parameters and investigate some properties of the solutions
such as existence of singularity(ies) and asymptotic behavior of them. Also,
we calculate the finite action, thermodynamic and conserved quantities of
the solutions and using the the Smarr-type formula to check the first law of
thermodynamics.
\end{abstract}

\maketitle

In recent years, one of the great developments in general relativity was the
discovery of a close relationship between black hole mechanics and the
ordinary laws of thermodynamics. The existence of this close relationship
between these laws may provide us with a key to our understanding of the
fundamental nature of black holes as well as to our understanding of some
aspects of the nature of thermodynamics itself. Also, it is notable that the
black hole is an object that is considered in classical and quantum point of
view and so one hopes to gain insight into the nature of quantum gravity by
studying the thermodynamics of black holes.

The laws of black hole mechanics are analogous to the laws of
thermodynamics. The quantities of particular interest in gravitational
thermodynamics are the physical entropy $S$ and the temperature $\beta ^{-1}$%
, where these quantities are respectively proportional to the area and
surface gravity of the event horizon \cite{Bek}. Other black hole
properties, such as energy, angular momentum and conserved charges can also
be given a thermodynamic interpretation. In finding the thermodynamic
quantities, one should use the quasilocal definitions for the thermodynamic
variables. By quasilocal, we mean that the quantity is constructed from
information that exists on the boundary of a gravitating system alone. Just
as the Gauss law, such quasilocal quantities will yield information about
the spacetime contained within the system boundary. One of the advantage of
using such a quasilocal method is that the formalism does not depend on the
particular asymptotic behavior of the system, so one can accommodate a wide
class of spacetimes with the same formalism.

In this paper, we attempt to construct the rotating black brane solutions of
Gauss-Bonnet gravity in the presence of a nonlinear Maxwell field, namely,
power Maxwell invariant, and investigate their thermodynamics properties. In
what follows, at first, we present some considerable works on higher
derivative gravity as well as power Maxwell invariant theory.

On one hand, since the field equations of gravity are generally covariant
and of second order derivatives in the metric tensor, one would naively
expect these equations to be derived from an action principle involving
metric tensor and its first and second order derivatives \cite{Lov},
analogous to the situation for many other field theories of physics. In
recent years, there have been considerable works for understanding the role
of the higher curvature terms from various points of view, especially with
regard to higher dimensional black hole physics. For example, thermodynamics
and other properties of the static black hole solutions in Gauss-Bonnet
gravity have been found by many authors \cite{Des,Wil1,MS,Cai,Ish,Wil11,Od11}
Also, the Taub-NUT/bolt solutions of higher derivative gravity and their
thermodynamics properties have been constructed \cite{DM2,Hendi,
Khoddam,HendiNUT}. Not long ago, M. H. Dehghani introduced two new classes
of rotating solutions of second order Lovelock gravity and investigated
their thermodynamics \cite{Deh1, Deh2}.

On the other hand, in recent years there has been aroused interest about the
solutions whose source is Maxwell invariant raised to the power $s$, i.e., $%
(F^{\mu \nu }F_{\mu \nu })^{s}$ as the source of geometry in Einstein and
higher derivative gravity \cite{Martinez1}. This theory is considerably
richer than that of the linear electromagnetic field and in the special case
$(s=1)$ it can reduces to linear field. Also, it is valuable to find and
analyze the effects of exponent $s$ on the behavior of the new solutions and
the laws of black hole mechanics \cite{Rasheed}. In addition, in higher
dimensional gravity, for the special choice $s=d/4$, where $d=$ dimension of
the spacetime is a multiple of $4$, it yields a traceless Maxwell's
energy-momentum tensor which leads to conformal invariance \cite{Martinez2}.
In Ref. \cite{Radu}, higher dimensional, direct analogues of the usual $d=4$
Einstein--Yang-Mills gravity have been studied.

In the rest of the paper, we give a brief definition of the field equations
of Gauss-Bonnet gravity in the presence of nonlinear electromagnetic field
and present a new class of rotating black brane solutions and investigate
their properties.

Here we present the Gauss-Bonnet gravity, which contains the first three
terms of Lovelock gravity in the presence of nonlinear electromagnetic
field. The action is
\begin{eqnarray}
I_{G} &=&-\frac{1}{16\pi }\int_{\mathcal{M}}d^{n+1}x\sqrt{-g}\left\{
R-2\Lambda +\alpha L_{GB}+\kappa L(F)\right\}  \nonumber \\
&&-\frac{1}{8\pi }\int_{\partial \mathcal{M}}d^{n}x\sqrt{-\gamma }\left\{
K+2\alpha \left( J-2\widehat{G}_{ab}K^{ab}\right) \right\} ,  \label{IG}
\end{eqnarray}
where $R$ is the Ricci scalar, $\Lambda $ is the cosmological constant, $%
\alpha $ is the Gauss-Bonnet coefficient with dimension $(\mathrm{length}
)^{2}$and $L_{GB}$ is the Lagrangian of Gauss-Bonnet gravity
\[
L_{GB}=R_{\mu \nu \gamma \delta }R^{\mu \nu \gamma \delta }-4R_{\mu \nu
}R^{\mu \nu }+R^{2},
\]
where $R_{\mu \nu }$ and $R_{\mu \nu \gamma \delta }$\ are Ricci and Riemann
tensors of the manifold $\mathcal{M}$, $\kappa $ is a constant in which one
can set it to avoid of negative energy density and present well-defined
solutions and $L(F)$ is the Lagrangian of power Maxwell invariant theory
\begin{equation}
L(F)=-F^{s}.  \label{ActEB}
\end{equation}
In Eq. (\ref{ActEB}), $F=F^{\mu \nu }F_{\mu \nu }$ where $F_{\mu \nu
}=\partial _{\mu }A_{\nu }-\partial _{\nu }A_{\mu }$\ is electromagnetic
tensor field and $A_{\mu }$\ is the vector potential. In the limit $s=1$, $%
L(F)$ reduces to the standard Maxwell form $L(F)=-F$. The second integral in
Eq. (\ref{IG}) is a boundary term which is chosen such that the variational
principle is well defined \cite{MyeDavis}. In this term, $\gamma _{ab}$ is
induced metric on the boundary $\partial \mathcal{M}$ , $K$ is trace of
extrinsic curvature $K^{ab}$ of the boundary, $\widehat{G}^{ab}(\gamma )$ is
Einstein tensor calculated on the boundary, and $J$ is trace of
\begin{equation}
J_{ab}=\frac{1}{3}
(K_{cd}K^{cd}K_{ab}+2KK_{ac}K_{b}^{c}-2K_{ac}K^{cd}K_{db}-K^{2}K_{ab}).
\label{psi}
\end{equation}

Varying the action (\ref{IG}) with respect to the metric tensor $g_{\mu \nu}$
and electromagnetic field $A_{\mu}$, the equations of gravitation and
electromagnetic fields are obtained as
\begin{eqnarray}
&&G_{\mu \nu }+\Lambda g_{\mu \nu }-\frac{\alpha}{2} \left( 8R^{\rho \sigma
}R_{\mu \rho \nu \sigma }-4R_{\mu }^{\ \rho \sigma \lambda }R_{\nu \rho
\sigma \lambda }-4RR_{\mu \nu }+\right.  \nonumber \\
&& \left. 8R_{\mu \lambda }R_{\text{ \ }\nu }^{\lambda }+ g_{\mu \nu
}L_{GB}\right) =2\kappa \left( sF_{\mu \rho }F_{\nu }^{~\rho }F^{s-1}-\frac{1%
}{4}g_{\mu \nu }F^{s}\right) ,  \label{Geq}
\end{eqnarray}
\begin{equation}
\partial _{\mu }\left( \sqrt{-g}F^{\mu \nu }F^{s-1}\right) =0,  \label{Maxeq}
\end{equation}
where $G_{\mu \nu }$ is the Einstein tensor.

Equation (\ref{Geq}) does not contain the derivative of the curvatures and
therefore the derivatives of the metric higher than two do not appear. Thus,
the Gauss-Bonnet gravity is a special case of higher derivative gravity.
Hereafter we consider the metric of $(n+1)$-dimensional rotating spacetime
with $k$ rotation parameters in the form \cite{Lemos,Awad}
\begin{eqnarray}
ds^{2} &=&-f(\rho )\left( \Xi dt-{{\sum_{i=1}^{k}}}a_{i}d\phi _{i}\right)
^{2}+\frac{\rho ^{2}}{l^{4}}{{\sum_{i=1}^{k}}}\left( a_{i}dt-\Xi l^{2}d\phi
_{i}\right) ^{2}  \nonumber \\
&&\ \text{ }+\frac{d\rho ^{2}}{f(\rho )}-\frac{\rho ^{2}}{l^{2}}{%
\sum_{i<j}^{k}}(a_{i}d\phi _{j}-a_{j}d\phi _{i})^{2}+\rho ^{2}dX^{2},
\label{met1}
\end{eqnarray}%
where $\Xi =\sqrt{1+\sum_{i}^{k}a_{i}^{2}/l^{2}}$ and $dX^{2}$ is the
Euclidean metric on the $\left( n-1-k\right) $-dimensional submanifold. The
rotation group in $(n+1)$ dimensions is $SO(n)$ and therefore $k\leq \lbrack
n/2]$. Using the gauge potential ansatz
\begin{equation}
A_{\mu }=h(\rho )\left( \Xi \delta _{\mu }^{0}-\delta _{\mu
}^{i}a_{i}\right) \text{(no sum on }i\text{)}  \label{Amu}
\end{equation}%
and solving Eq. (\ref{Maxeq}), we obtain
\begin{equation}
h(\rho )=\left\{
\begin{array}{cc}
C, & s=0,\frac{1}{2} \\
C\ln (\rho ), & s=\frac{n}{2} \\
C\rho ^{(2s-n)/(2s-1)}, & \text{otherwise}%
\end{array}%
\right. ,
\end{equation}%
where $C$ is an integration constant which is related to the charge
parameter. One may note that as $s=1$, $A_{\mu }$ of Eq. (\ref{Amu}) reduces
to the gauge potential of the linear Maxwell field \cite{Deh1} as it should
be. To find the function $f(\rho )$ , one may use any components of Eq. (\ref%
{Geq}). After some calculations, one can show that the solution of field
equation (\ref{Geq}), can be written as
\begin{equation}
f(\rho )=\frac{2\rho ^{2}}{(n-1)\gamma }\left( 1-\sqrt{1+\frac{2\gamma
\Lambda }{n}+\frac{\gamma m}{\rho ^{n}}-\kappa \gamma \Gamma (\rho )}\right)
,  \label{fr}
\end{equation}%
where
\begin{eqnarray}
\Gamma (\rho ) &=&\left\{
\begin{array}{cc}
0, & s=0,\frac{1}{2} \\
2^{n/2}(n-1)C^{n}\frac{\ln (\rho )}{\rho ^{n}}, & s=\frac{n}{2} \\
\frac{(2s-1)^{2}}{2s-n}\left( \frac{-2C^{2}\rho ^{-2(n-1)/(2s-1)}(2s-n)^{2}}{%
(2s-1)^{2}}\right) ^{s}, & \text{Otherwise}%
\end{array}%
\right. ,  \label{Gamma} \\
\gamma  &=&\frac{4\alpha (n-2)(n-3)}{(n-1)}.  \label{gamma}
\end{eqnarray}%
The metric function $f(\rho )$ for the uncharged solution ($C=0$) is real in
the whole range $0\leq \rho <\infty $ provided $\alpha \leq -n/\left(
8\gamma \Lambda \right) $, but for charged solution it is real only in the
range $r_{0}\leq \rho <\infty $\ where $r_{0}$\ is the largest real root of
the following equation
\begin{equation}
nr_{0}^{n}+\gamma \left( 2\Lambda r_{0}^{n}+nm-\kappa nr_{0}^{n}\Gamma
_{0}\right) =0,  \label{r0Eq}
\end{equation}%
where $\Gamma _{0}=\Gamma (\rho =r_{0})$. In order to restrict the spacetime
to the region $\rho \geq r_{0}$, we introduce a new radial coordinate $r$ as
\begin{equation}
r=\sqrt{\rho ^{2}-r_{0}^{2}}\Rightarrow d\rho ^{2}=\frac{r^{2}}{%
r^{2}+r_{0}^{2}}dr^{2}.  \label{transformation}
\end{equation}%
With this new coordinate, the above metric (\ref{met1}) becomes
\begin{eqnarray}
ds^{2} &=&-f(r)\left( \Xi dt-{{\sum_{i=1}^{k}}}a_{i}d\phi _{i}\right) ^{2}+%
\frac{(r^{2}+r_{0}^{2})}{l^{4}}{{\sum_{i=1}^{k}}}\left( a_{i}dt-\Xi
l^{2}d\phi _{i}\right) ^{2}  \nonumber \\
&&-\frac{r^{2}+r_{0}^{2}}{l^{2}}{\sum_{i<j}^{k}}(a_{i}d\phi _{j}-a_{j}d\phi
_{i})^{2}+\frac{r^{2}dr^{2}}{(r^{2}+r_{0}^{2})f(r)}+(r^{2}+r_{0}^{2})dX^{2},
\label{met2}
\end{eqnarray}%
where now the functions $h(\rho )$ and $f(\rho )$ change to
\begin{eqnarray}
&&h(r)=\left\{
\begin{array}{cc}
C, & s=0,\frac{1}{2} \\
\frac{1}{2}C\ln (r^{2}+r_{0}^{2}), & s=\frac{n}{2} \\
C(r^{2}+r_{0}^{2})^{(2s-n)/[2(2s-1)]}, & \text{otherwise}%
\end{array}%
\right. ,  \label{A2} \\
&&f(r)=\frac{2\left( r^{2}+r_{0}^{2}\right) }{(n-1)\gamma }\left( 1-\sqrt{1+%
\frac{2\gamma \Lambda }{n}+\frac{\gamma m}{(r^{2}+r_{0}^{2})^{n/2}}-\kappa
\gamma \Gamma (r)}\right) ,  \label{fr2} \\
\Gamma (r) &=&\left\{
\begin{array}{cc}
0, & s=0,\frac{1}{2} \\
2^{n/2}(n-1)C^{n}\frac{\ln (r^{2}+r_{0}^{2})}{2(r^{2}+r_{0}^{2})^{n/2}}, & s=%
\frac{n}{2} \\
\frac{(2s-1)^{2}}{2s-n}\left( \frac{-2C^{2}(2s-n)^{2}}{%
(2s-1)^{2}(r^{2}+r_{0}^{2})^{(n-1)/(2s-1)}}\right) ^{s}, & \text{Otherwise}%
\end{array}%
\right. ,  \label{gr2}
\end{eqnarray}%
These solutions reduce to the solutions presented in Ref. \cite{Deh1} as $s=1
$ , to those of Ref. \cite{HendiEnon} as $\alpha $ vanishes and to the
solutions given in Ref. \cite{DehEMax} as $\alpha $ goes to zero and $s=1$,
simultaneously.

{\Large Properties of the solutions}

In order to study the general structure of these spacetime, we first look
for the essential singularity(ies). After some algebraic manipulation, one
can show that for the rotating metric (\ref{met2}), the Kretschmann scalar
is
\begin{eqnarray}
R_{\mu \nu \rho \sigma }R^{\mu \nu \rho \sigma } &=&\frac{%
(r^{2}+r_{0}^{2})^{2}f^{\prime \prime 2}(r)}{r^{4}}-\frac{%
2r_{0}^{2}(r^{2}+r_{0}^{2})f^{\prime \prime }(r)f^{\prime }(r)}{r^{5}}+
\nonumber \\
&&\frac{\lbrack 2(n-1)r^{4}+r_{0}^{4})]f^{\prime 2}(r)}{r^{6}}+\frac{%
2(n-1)(n-2)f^{2}(r)}{(r^{2}+r_{0}^{2})^{2}},  \label{RR}
\end{eqnarray}%
where prime and double prime are first and second derivative with respect to
$r$ , respectively. It is straightforward to show that the Kretschmann
scalar (\ref{RR}) with metric function (\ref{fr2}) diverges at $r=0$ and is
finite for $r\neq 0$. Also one can show that other curvature invariants
(such as Ricci square, Weyl square and so on) diverges at $r=0$. Thus, there
is a curvature singularity located at $r=0$.

In Einstein gravity coupled to power Maxwell invariant \cite{HendiEnon}, it
is shown that for $s>n/2$ and $s<0$, the singularity at $r=0$ for the
solutions with non-negative mass is spacelike, and therefore it is
unavoidable. These solutions with positive mass present black branes with
one horizon. But here, in Gauss-Bonnet gravity, we have a extra parameter, $%
\alpha $, and one can set the parameters of the solutions to have timelike
singularity for all values of nonlinear parameter.

Second, we investigate the effects of the nonlinearity on the asymptotic
behavior of the solutions. It is worthwhile to mention that for $0<s<\frac{1
}{2}$, the asymptotic dominant term of Eq. (\ref{fr2}) is charge term, $%
\Gamma(r)$, and the presented solutions are not asymptotically AdS, but for
the cases $s<0$ or $s>\frac{1}{2}$ (include of $s=\frac{n}{2}$), the
asymptotic behavior of rotating Einstein-nonlinear Maxwell field solutions
are the same as linear AdS case. It is easy to show that the electromagnetic
field is zero for the cases $s=0,1/2$, and the metric function (\ref{fr2})
does not possess a charge term ($\Gamma(r)=0$) and it corresponds to
uncharged asymptotically AdS one.

The metric (\ref{met2}) has two types of Killing and event horizons. The
Killing horizon is a null surface whose null generators are tangent to a
Killing field. It is proved that a stationary black hole event horizon
should be a Killing horizon in the four-dimensional Einstein gravity \cite%
{Haw1}. This proof can not obviously be generalized to higher order gravity,
but the result is true for all the known static solutions. Although our
solution is not static, the Killing vector,
\begin{equation}
\chi =\partial _{t}+{\sum_{i}^{k}}\Omega _{i}\partial _{\phi _{i}},
\label{Kil}
\end{equation}
is the null generator of the event horizon, where $\Omega _{i}$ is the $i$th
component of angular velocity of the outer horizon. The angular velocities $%
\Omega _{i}$'s may be obtained by analytic continuation of the metric.
Setting $a_{i}\rightarrow ia_{i}$ yields the Euclidean section of (\ref{met2}%
), whose regularity at $r=r_{+}$ (requiring the absence of conical
singularity at the horizon in the Euclidean sector of the black brane
solutions) requires that we should identify $\phi _{i}\sim \phi _{i}+\beta
\Omega _{i}$. One obtains
\begin{equation}
\Omega _{i}=\frac{a_{i}}{\Xi l^{2}}.  \label{Om}
\end{equation}
The temperature may be obtained through the use of regularity at $r=r_{+}$
or definition of surface gravity,
\begin{equation}
T_{+}{=\beta }^{-1}=\frac{1}{2\pi }\sqrt{-\frac{1}{2}\left( \nabla _{\mu
}\chi _{\nu }\right) \left( \nabla ^{\mu }\chi ^{\nu }\right) ,}
\end{equation}
where $\chi $ is the Killing vector (\ref{Kil}). It is straightforward to
show that
\begin{eqnarray}
{T}_{+} &{=}&\frac{f^{\prime }(r_{+})}{4\pi \Xi }\sqrt{1+\frac{r_{0}^{2}}{
r_{+}^{2}}}=-\frac{2\Lambda +(n-1)\kappa \Upsilon }{4\pi \Xi (n-1)}\sqrt{
r_{+}^{2}+r_{0}^{2}},  \label{Temp} \\
\Upsilon &=&\left\{
\begin{array}{cc}
0, & s=0,\frac{1}{2} \\
2^{n/2}C^{n}(r_{+}^{2}+r_{0}^{2})^{-n/2}, & s=\frac{n}{2} \\
\frac{(2s-1)}{n-1}\left( \frac{-2C^{2}(2s-n)^{2}}{
(2s-1)^{2}(r_{+}^{2}+r_{0}^{2})^{(n-1)/(2s-1)}}\right) ^{s}, & \text{
otherwise}%
\end{array}
\right. .
\end{eqnarray}
Using the fact that the temperature of the extreme black brane is zero, it
is easy to show that the condition for having an extreme black hole is that
the mass parameter is equal to $m_{\mathrm{ext}}$, where $m_{\mathrm{ext}}$
is given as
\begin{equation}
m_{\mathrm{ext}}=\zeta -\frac{8\Lambda (r^{2}+r_{0}^{2})^{n/2}}{n(n-4)}+
\frac{(r^{2}+r_{0}^{2})^{n/2}\left[ \sqrt{(n-1)(n-2)\xi }-(n-1)(n-2)\right]
}{\alpha (n-2)(n-3)(n-4)^{2}}  \label{mext}
\end{equation}
where%
\begin{eqnarray*}
\chi &=&\left\{
\begin{array}{cc}
2^{n/2}\kappa (n-1)C^{n}(r^{2}+r_{0}^{2})^{-n/2}, & s=\frac{n}{2} \\
(2s-1)\kappa \left( \frac{-2C^{2}(n-2s)^{2}}{
(2s-1)^{2}(r^{2}+r_{0}^{2})^{(n-1)/(2s-1)}}\right) ^{s}, & \text{otherwise}%
\end{array}
\right. , \\
\zeta &=&\left\{
\begin{array}{cc}
\frac{2^{n/2}\kappa (n-1)C^{n}[(n-4)\ln (r^{2}+r_{0}^{2})-2]}{2(n-4)}, & s=
\frac{n}{2} \\
-\frac{2\kappa (2s-1)\left[ s(n-5)+2\right] (r^{2}+r_{0}^{2})^{n/2}}{
(n-4)(n-2s)}\left( \frac{-2C^{2}(n-2s)^{2}}{
(2s-1)^{2}(r^{2}+r_{0}^{2})^{(n-1)/(2s-1)}}\right) ^{s}, & \text{otherwise}%
\end{array}
\right. , \\
\xi &=&(n-1)(n-2)+8\alpha \Lambda (n-3)(n-4)+4(n-3)(n-4)\alpha \chi .
\end{eqnarray*}
The metric of Eqs. (\ref{met2}), (\ref{fr2}) and (\ref{gr2}) presents a
black brane solution with inner and outer horizons, provided the mass
parameter $m$ is greater than $m_{\mathrm{ext}}$, an extreme black brane for
$m=m_{\mathrm{ext}}$ and a naked singularity otherwise.

Here we, first, calculate the thermodynamic and conserved quantities of the
black brane. Second, we obtain a Smarr-type formula for the mass as a
function of the entropy, the angular momentum and the charge of the solution
and finally check the first law of thermodynamics.

Denoting the volume of the hypersurface at $r=$constant and $t=$constant by $%
V_{n-1}$, the charge of the black brane, $Q$, can be found by calculating
the flux of the electromagnetic field at infinity, yielding
\begin{equation}
Q=\left\{
\begin{array}{cc}
\frac{V_{n-1}(-1)^{(n+2)/2}2^{n/2}\Xi C^{n-1}}{8\pi }, & s=\frac{n}{2} \\
\frac{V_{n-1}(-1)^{s+1}2^{s}\Xi }{8\pi }\left( \frac{(2s-n)C}{(2s-1)}\right)
^{2s-1}, & \text{otherwise}%
\end{array}
\right.  \label{Charg}
\end{equation}
The electric potential $\Phi$, measured at infinity with respect to the
horizon, is defined by \cite{Gub}
\begin{equation}
\Phi =A_{\mu }\chi ^{\mu }\left\vert _{r\rightarrow \infty }-A_{\mu }\chi
^{\mu }\right\vert _{r=r_{+}},  \label{Pot}
\end{equation}
where $\chi $ is the null generator of the horizon given by Eq. (\ref{Kil}).
One finds
\begin{equation}
\Phi =\frac{-C}{\Xi }\left\{
\begin{array}{cc}
1, & s=0,\frac{1}{2} \\
\frac{1}{2}\ln (r_{+}^{2}+r_{0}^{2}), & s=\frac{n}{2} \\
(r_{+}^{2}+r_{0}^{2})^{(2s-n)/[2(2s-1)]}, & \text{otherwise}%
\end{array}
\right. .
\end{equation}

Black hole entropy typically satisfies the so-called area law, which states
that the entropy of a black hole equals one-quarter of the area of its
horizon. This near universal law applies to almost all kinds of black
objects in Einstein gravity \cite{Beck}. However in higher derivative
gravity the area law is not satisfied in general \cite{fails}. For
asymptotically flat black hole solutions of Lovelock gravity, the entropy
may be written as \cite{Myers2}
\begin{equation}
S=\frac{1}{4}\sum_{k=1}^{[(d-1)/2]}k\alpha _{k}\int d^{n-1}x\sqrt{\tilde{g}}
\tilde{\mathcal{L}}_{k-1},  \label{Enta}
\end{equation}%
where the integration is done on the $(n-1)$-dimensional spacelike
hypersurface of Killing horizon, $\tilde{g}_{\mu \nu }$ is the induced
metric on it, $\tilde{g}$ is the determinant of $\tilde{g}_{\mu \nu }$ and $%
\tilde{\mathcal{L}}_{k}$ is the $k$th order Lovelock Lagrangian of $\tilde{g}%
_{\mu \nu }$. The asymptotic behavior of the black branes we are considering
is not flat, and therefore we calculate the entropy through the use of
Gibbs-Duhem relation
\begin{equation}
S=\frac{1}{T}(\mathcal{M}-\Gamma _{i}\mathcal{C}_{i})-I,  \label{GibsDuh}
\end{equation}%
where $I$ is the finite total action evaluated on the classical solution,
and $\mathcal{C}_{i}$ and $\Gamma _{i}$ are the conserved charges and their
associate chemical potentials respectively.

In general the action $I_{G}$, is divergent when evaluated on the solutions,
as is the Hamiltonian and other associated conserved quantities. A
systematic method of dealing with this divergence in Einstein gravity is
through the use of the counterterms method inspired by the anti-de
Sitter/conformal field theory (AdS/CFT) correspondence \cite{Mal}. This
conjecture, which relates the low energy limit of string theory in
asymptotically anti de-Sitter spacetime and the quantum field theory living
on the boundary of it, have attracted a great deal of attention in recent
years. This equivalence between the two formulations means that, at least in
principle, one can obtain complete information on one side of the duality by
performing computation on the other side. A dictionary translating between
different quantities in the bulk gravity theory and their counterparts on
the boundary has emerged, including the partition functions of both
theories. This conjecture is now a fundamental concept that furnishes a
means for calculating the action and conserved quantities intrinsically
without reliance on any reference spacetime \cite{Sken1BKOd1}. It has also
been applied to the case of black holes with constant negative or zero
curvature horizons \cite{DehEMax} and rotating higher genus black branes
\cite{Deh4}. Although the AdS/CFT correspondence applies for the case of a
specially infinite boundary, it was also employed for the computation of the
conserved and thermodynamic quantities in the case of a finite boundary \cite%
{Deh5}.

All of the work mention in the last paragraph was limited to Einstein
gravity. Although the counterterms in Lovelock gravity should be a scalar
constructed from Riemann tensor as in the case of Einstein gravity, they are
not known for the case of Lovelock gravity till now. But, for the solutions
with flat boundary, $\widehat{R}_{abcd}(\gamma )=0$, there exists only one
boundary counterterm
\begin{equation}
I_{\mathrm{ct}}=-\frac{1}{8\pi }\int_{\partial \mathcal{M}}d^{n}x\sqrt{%
-\gamma }\left( \frac{n-1}{l_{\mathrm{eff}}}\right) ,  \label{Ict}
\end{equation}%
where $l_{\mathrm{eff}}$\ is a scale length factor that depends on $l$ and $%
\alpha $, that must reduce to $l$ as $\alpha $ goes to zero. One may note
that this counterterm has exactly the same form as the counterterm in
Einstein gravity for a spacetime with zero curvature boundary in which $l$
is replaced by $l_{\mathrm{eff}}$.

Using Eqs. (\ref{IG}) and (\ref{Ict}), the finite action can be calculated
as
\begin{eqnarray}
I &=&\frac{V_{n-1}\Xi l^{2}(r_{+}^{2}+r_{0}^{2})^{(n-1)/2}}{4}\left( \Pi -
\frac{1}{l^{2}}\right) ,  \label{finiteACT} \\
\Pi &=&\left\{
\begin{array}{cc}
\frac{\left( n-1\right) m}{n}(r_{+}^{2}+r_{0}^{2})^{-n/2}, & s=0,\frac{1}{2}
\\
\frac{\left( n-1\right) m+(-1)^{(n+2)/2}2^{n/2}C^{n}\ln
(r_{+}^{2}+r_{0}^{2}) }{n-2^{n/2}l^{2}\kappa C^{n}}, & s=\frac{n}{2} \\
\frac{\left( n-1\right)
^{2}(2s-1)^{2s-1}m(r_{+}^{2}+r_{0}^{2})^{(n-2s)/[2(2s-1)]}-(-1)^{s}2^{s+1}(n-1)(2s-n)^{2s-1}C^{2s}
}{n(n-1)(2s-1)^{2s-1}-(-1)^{s}2^{s}l^{2}\kappa (2s-n)^{2s}C^{2s}}, & \text{
otherwise}%
\end{array}
\right. ,
\end{eqnarray}
Having the total finite action, $I_{G}+I_{\mathrm{ct}}$, one can use the
Brown-York definition of stress energy tensor \cite{Brown} to construct a
divergence-free stress energy tensor. For the case of manifolds with zero
curvature boundary, the finite stress energy tensor is \cite{DM1,DAH}
\begin{equation}
T^{ab}=\frac{1}{8\pi }\{(K^{ab}-K\gamma ^{ab})+2\alpha (3J^{ab}-J\gamma
^{ab})-\left( \frac{n-1}{l_{\mathrm{eff}}}\right) \gamma ^{ab}\}
\label{Stres}
\end{equation}
One may note that when $\alpha $ goes to zero, the stress energy tensor (\ref%
{Stres}) reduces to that of Einstein gravity. To compute the conserved
charges of the spacetime, we choose a spacelike surface $\mathcal{B}$ in $%
\partial \mathcal{M}$ with metric $\sigma _{ij}$, and write the boundary
metric in ADM form
\begin{equation}
\gamma _{ab}dx^{a}dx^{a}=-N^{2}dt^{2}+\sigma _{ij}\left( d\varphi
^{i}+V^{i}dt\right) \left( d\varphi ^{j}+V^{j}dt\right) ,
\end{equation}
where the coordinates $\varphi ^{i}$ are the angular variables
parameterizing the hypersurface of constant $r$ around the origin, and $N$
and $V^{i}$ are the lapse and shift functions respectively. When there is a
Killing vector field $\mathcal{\xi }$ on the boundary, then the quasilocal
conserved quantities associated with the stress energy tensors of Eq. (\ref%
{Stres}) can be written as
\begin{equation}
\mathcal{Q}(\mathcal{\xi )}=\int_{\mathcal{B}}d^{n-1}\varphi \sqrt{\sigma }
T_{ab}n^{a}\mathcal{\xi }^{b},  \label{charge}
\end{equation}
where $\sigma $ is the determinant of the metric $\sigma _{ij}$, and $n^{a}$
is the timelike unit normal vector to the boundary $\mathcal{B}$. For
boundaries with timelike ($\xi =\partial /\partial t$) and rotational ($%
\varsigma =\partial /\partial \varphi $) Killing vector fields, one obtains
the quasilocal mass and angular momentum
\begin{eqnarray}
M &=&\int_{\mathcal{B}}d^{n-1}\varphi \sqrt{\sigma }T_{ab}n^{a}\xi ^{b},
\label{Mas} \\
J &=&\int_{\mathcal{B}}d^{n-1}\varphi \sqrt{\sigma }T_{ab}n^{a}\varsigma
^{b},  \label{Amom}
\end{eqnarray}
provided the surface $\mathcal{B}$ contains the orbits of $\varsigma $.
These quantities are, respectively, the conserved mass and angular momentum
of the system enclosed by the boundary $\mathcal{B}$. Using Eqs. (\ref{Mas})
and (\ref{Amom}), the mass and angular momenta of the solution are
calculated as
\begin{eqnarray}
M &=&\frac{V_{n-1}}{16\pi }m\left( n\Xi ^{2}-1\right) ,  \label{Mass} \\
J_{i} &=&\frac{V_{n-1}}{16\pi }n\Xi ma_{i}.  \label{Angmom}
\end{eqnarray}

Now using Gibbs-Duhem relation (\ref{GibsDuh}) and Eqs. (\ref{Charg}) - (\ref%
{Angmom}) and (\ref{finiteACT}), one obtains
\begin{equation}
S=\frac{V_{n-1}\Xi }{4}(r_{+}^{2}+r_{0}^{2})^{(n-1)/2}.  \label{Entropy}
\end{equation}
This shows that the entropy obeys the area law for our case where the
horizon curvature is zero.

Calculating all the thermodynamic and conserved quantities of the black
brane solutions, we now check the first law of thermodynamics for our
solutions. We obtain the mass as a function of the extensive quantities $S$,
$\mathbf{J}$, and $Q$. Using the expression for the mass, the angular
momenta, the charge and the entropy given in Eqs. (\ref{Charg}), (\ref{Mass}%
), (\ref{Angmom}), (\ref{Entropy}) and the fact that $f(r_{+})=0$, one can
obtain a Smarr-type formula as
\begin{equation}
M(S,\mathbf{J},Q)=\frac{(nZ-1)J}{nl\sqrt{Z(Z-1)}},  \label{Smar}
\end{equation}%
where $J=\left\vert \mathbf{J}\right\vert =\sqrt{\sum_{i}^{k}J_{i}^{2}}$ and
$Z=\Xi ^{2}$ is the positive real root of the following equation

\begin{equation}
2\Lambda +\frac{16\pi J}{lZ\sqrt{Z^{2}-1}}\left( \frac{Z}{4S}\right)
^{n/(n-1)}-\kappa n\Psi =0,  \label{Zsmar}
\end{equation}%
\[
\Psi =\left\{
\begin{array}{cc}
0, & s=0,\frac{1}{2} \\
2^{n/[2(n-1)]}\left[ \frac{(-1)^{(n+2)/2}\pi Q}{S}\right] ^{n/(n-1)}\ln
\left( \frac{4S}{Z}\right) , & s=\frac{n}{2} \\
\frac{(-1)^{s}2^{s/(2s-1)}(2s-1)^{2}}{2s-n}\left( \frac{(-1)^{s+1}\pi Q}{S}%
\right) ^{2s/(2s-1)}, & \text{Otherwise}%
\end{array}%
\right. .
\]%
One may then regard the parameters $S$, $J_{i}$'s, and $Q$ as a complete set
of extensive parameters for the mass $M(S,\mathbf{J},Q)$ and define the
intensive parameters conjugate to them. These quantities are the
temperature, the angular velocities, and the electric potential
\begin{equation}
T=\left( \frac{\partial M}{\partial S}\right) _{J,Q},\ \ \Omega _{i}=\left(
\frac{\partial M}{\partial J_{i}}\right) _{S,Q},\ \ \Phi =\left( \frac{%
\partial M}{\partial Q}\right) _{S,J}  \label{Dsmar}
\end{equation}%
It is a matter of straightforward calculation to show that the intensive
quantities calculated by Eq. (\ref{Dsmar}) coincide with Eqs. (\ref{Om}), (%
\ref{Temp}), and (\ref{Pot}). Thus, these quantities satisfy the first law
of thermodynamics
\begin{equation}
dM=TdS+{{{\sum_{i=1}^{k}}}}\Omega _{i}dJ_{i}+\Phi dQ.  \label{FirstLaw}
\end{equation}

\section{ CLOSING REMARKS}

In this paper, we presented a class of rotating solutions in Gauss-Bonnet
gravity in the presence of a nonlinear electromagnetic field. These
solutions are not real for the whole spacetime and so, by a suitable
transformation, we presented the real solutions. We found that these
solutions reduce to the solutions of Gauss--Bonnet--Maxwell gravity as $s=1$
, and reduce to those of Einstein-power Maxwell invariant gravity as $\alpha$
vanishes. Then we studied the kind of singularity and found that, in
contrast to the Einstein-power Maxwell invariant gravity, for all values of
nonlinear parameter, we can always choose the suitable parameters to have
timelike singularity. Also, we found that as in the case of rotating black
brane solutions of Einstein-power Maxwell invariant gravity, for $0<s<\frac{%
1 }{2}$, the asymptotic dominant term of metric function $f(r)$ is charge
term, and the presented solutions are not asymptotically AdS, but for the
cases $s<0$ or $s>\frac{1}{2}$, the asymptotic behavior of rotating
Einstein-nonlinear Maxwell field solutions are the same as linear AdS case.
In the other word, we found that the Gauss-Bonnet does not effect on the
asymptotic behavior of the solutions. Then, we applied counterterm method to
the solutions with flat boundary at $r=$\textrm{constant} and $t=$\textrm{\
constant}, and calculated the finite action and their conserved and
thermodynamic quantities. The physical properties of the black brane such as
the temperature, the angular velocity, the electric charge and the potential
have been computed. We found that the conserved quantities of the black
brane do not depend on the Gauss-Bonnet parameter $\alpha $. Consequently,
we obtained the entropy of the black brane through the use of Gibbs-Duhem
relation and found that it obeys the area law of entropy. Then, we obtained
a Smarr-type formula for the mass as a function of the extensive parameters $%
S$ , $\mathbf{J}$ and $Q$, and calculated the intensive parameters $T$, $%
\Omega $ and $\Phi $. We also showed that the conserved and thermodynamic
quantities satisfy the first law of thermodynamics.

\begin{acknowledgements}
This work has been supported financially by Research Institute for
Astronomy and Astrophysics of Maragha.
\end{acknowledgements}

\end{document}